\documentclass[twocolumn,showpacs,preprintnumbers,amsmath,amssymb]{revtex4}

\usepackage{graphicx}% Include figure files
\usepackage{dcolumn}% Align table columns on decimal point
\usepackage{bm}% bold math

\begin{document}
\title{Temperature dependence of low-energy phonons in magnetic non-superconducting TbNi$_{2}$B$_{2}$C}
\author{S.~Anissimova$^{1}$, A.~Kreyssig$^{2,6}$, O.~Stockert$^{4}$, M.~Loewenhaupt$^{2}$, D.~Reznik$^{1,3,5}$}
\affiliation{$^{1}$Department of Physics, University of Colorado at Boulder, Boulder, Colorado 80309\\
$^{2}$Technische Universit$\ddot{a}$t Dresden, Institut f$\ddot{u}$r Festk$\ddot{o}$rperphysik, D-01062 Dresden, Germany\\
$^{3}$CEA Saclay, Laboratoire L$\acute{e}$on Brillouin, F-91191 Gif sur Yvette, France\\
$^{4}$Max-Planck-Institut f$\ddot{u}$r Chemische Physik fester Stoffe, N$\ddot{o}$thnitzer Stra$\beta$e 40, D-01187 Dresden, Germany\\
$^{5}$Karlsruher Institut f$\ddot{u}$r Technologie (KIT), Institute f$\ddot{u}$r Festk$\ddot{o}$rperphysik, Postfach 3640, D-76121 Karlsruhe, Germany\\
$^{6}$Ames Laboratory and Department of Physics and Astronomy, Iowa State University, Ames, Iowa 50011, USA}
\date{\today}
\begin{abstract}
We report temperature dependence of low-energy phonons in
magnetic nonsuperconducting TbNi$_{2}$B$_{2}$C single crystals
measured by inelastic neutron scattering. We observed 
low-temperature softening and broadening of two phonon branches,
qualitatively similar to that previously reported for
superconducting $\textit{R}$Ni$_{2}$B$_{2}$C (\textit{R} = rare earth, \textit{Y})
compounds. This result suggests that superconductivity in
TbNi$_{2}$B$_{2}$C compounds is absent not because of weak
electron-phonon coupling but as a result of
pair breaking due to magnetism.
\end{abstract}

\pacs{74.70.Dd,74.25.Kc,78.70.Nx, 63.20.kd\\
\\
\textit{Key words:} Borocarbides, Phonons, Neutron inelastic scattering, Electron-phonon interaction}

\maketitle

\section{INTRODUCTION}
\label{sec:INTRODUCTION}

In some rare-earth nickel borocarbides $\textit{R}$Ni$_{2}$B$_{2}$C, superconductivity 
coexists with magnetic order \cite{PhysRevB.50.647, PhysRevB.51.420}.
Extensive neutron and x-ray scattering experiments revealed an
incommensurate magnetic structure below approximately 15~K
in both superconducting Er, Ho
\cite{PhysRevB.51.678, PhysRevB.51.681, PhysRevB.50.9668,
PhysRevLett.75.2628}  and nonsuperconducting Tb, Gd
\cite{PhysRevB.53.8506, PhysRevB.53.6355} compounds. This observation
was interpreted in terms of common Fermi-surface nesting features along
\textbf{$a^{*}$},
which cause magnetic ordering of the rare-earth moments via the
Ruderman-Kittel-Kasuya-Yosida (RKKY) mechanism \cite{PhysRevB.53.8506}.
$^{57} Fe$ M$\ddot{o}$ssbauer spectroscopy and muon-
spin relaxation ($\mu$SR) studies of polycrystalline
TbNi$_{2}$B$_{2}$C \cite{PhysRevB.57.10268} confirmed
the presence of a small ferromagnetic component below about 8~K
previously observed via neutron diffraction \cite{PhysRevB.53.8506}
and magnetization measurements \cite{PhysRevB.53.307, PhysRevB.53.8499}.

In addition to magnetic effects, strong phonon softening has been observed in superconducting
$\textit{R}$Ni$_{2}$B$_{2}$C single crystals with $\textit{R} =$
Lu, Y, Er, and Ho \cite{PhysRevB.52.R9839, PhysRevB.55.R8678,
PhysRevLett.77.4628, PhysRevB.66.212503}, while no significant temperature dependence of
the phonon spectra was detected for the
nonsuperconducting TbNi$_{2}$B$_{2}$C \cite{kreyssig04}.
The superconducting transition temperature $T_c$
systematically decreases for $\textit{R}$Ni$_{2}$B$_{2}$C
($R$=Lu, Y, Tm, Er, and Ho) upon going from Lu ($T_c=$ 16.6~K)
to Ho ($T_c=$ 7.5~K). This observation was interpreted by H. Eisaki \textit{et al}. in terms of increasing coupling between the rare-earth magnetic moments and the conduction electrons \cite{PhysRevB.50.647}, which suppressed superconductivity. For TbNi$_{2}$B$_{2}$C, this pair breaking could be strong enough to completely destroy superconductivity. An alternative possibility for the absence of superconductivity in this system is that electron-phonon coupling is weaker than in the superconducting compounds \cite{kreyssig04}.

We investigated the strength of electron-phonon coupling in
TbNi$_{2}$B$_{2}$C by detailed measurements of the temperature
dependence of low-energy phonons in the magnetic
nonsuperconducting TbNi$_{2}$B$_{2}$C by inelastic neutron
scattering. The observed softening of these phonon branches in
this compound upon cooling from 300 to 30~K indicates that
electron-phonon interactions in magnetic nonsuperconducting
TbNi$_{2}$B$_{2}$C are strong and, although weaker, are of the
same order of magnitude as those in superconducting
$\textit{R}$Ni$_{2}$B$_{2}$C compounds.

\section{EXPERIMENTAL DETAILS}
\label{sec:EXPERIMENT}

\begin{figure}\vspace{-7mm}
\begin{center}
\scalebox{0.45}{\includegraphics{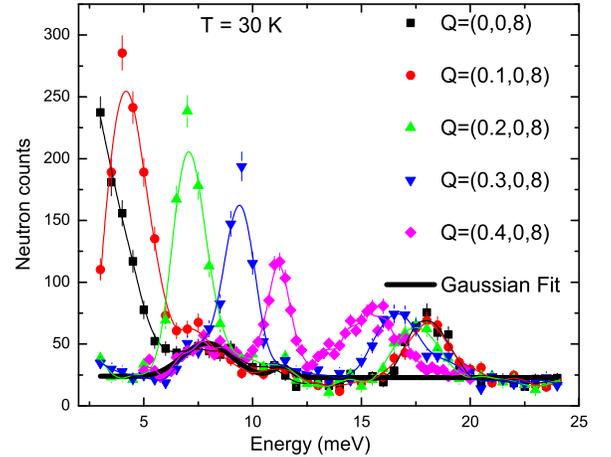}}\vspace{-1cm}
\end{center}
\caption{\label{fig1.} (Color online) The $\xi=0$, $\xi=0.1$, $\xi=0.2$,
$\xi=0.3$, and $\xi=0.4$ spectra at $T =$ 30~K. The bold solid
line represents the crystal electric-field excitations fit with a
Gaussian line shape. Solid lines represent smoothed data.}
\end{figure}

Intermetallic RNi$_{2}$B$_{2}$C borocarbides crystallize in the
body-centered tetragonal LuNi$_{2}$B$_{2}$C structure with the space
group $I4/mmm$, which consists of $R-$C layers separated by
Ni$_{2}$B$_{2}$ sheets \cite{siegrist94}. All crystallographic
parameters, i.e., the lattice contants $\textit{a}$ and
$\textit{c}$ as well as the $\textit{z}$ position of boron, scale
roughly with the ionic radius of the rare earth.
TbNi$_{2}$B$_{2}$C compounds exhibit antiferromagnetic ordering
at low temperatures below 15~K with the magnetic Tb$^{3+}$
moments aligned along the \textit{a} axis and modulated with
the propagation vector $\tau=(0.545,0,0)$ \cite{PhysRevB.53.8506,
PhysRevB.55.6584}. No superconductivity has been detected in this
compound down to 7~mK \cite{bitterlich98}.

Rodlike TbNi$_{2}$B$_{2}$C single crystals were grown by the
floating zone method using $^{11}$B isotope to avoid strong
neutron absorption \cite{souptel05}. Specimens with a length of
6~mm were cut from the rods with 6~mm diameter. For the inelastic
neutron-scattering experiments the samples were oriented in the
$\left(\textbf{a}-\textbf{c}\right)$ scattering plane. The measurements were
performed on the triple-axis spectrometer 1T1 at the Laboratoire
L$\acute{e}$on Brillouin, Saclay. Phonons propagating
in [$\xi$ ~0 ~0] direction, where $\xi$ is a reduced wave vector,
were recorded in the Brillouin zone centered at (0 ~0 ~8) by energy
scans using a fixed final energy $E_{f}=14.8$~meV at temperatures
between $T=$ 2 and 300~K.

\section{EXPERIMENTAL RESULTS}
\label{sec:EXPERIMENTALRESULTS}

\begin{figure}\vspace{-5mm}
\begin{center}
\scalebox{0.44}{\includegraphics{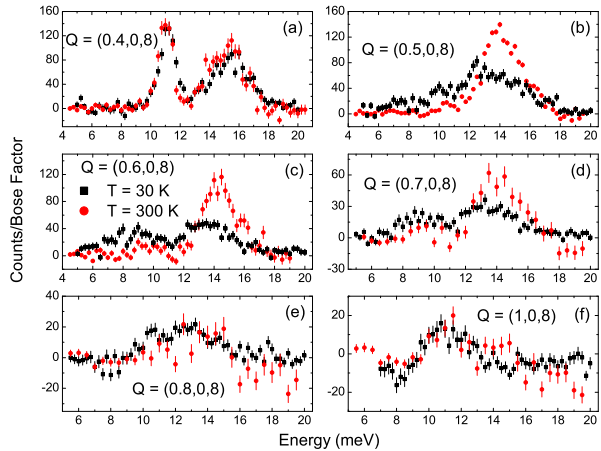}}\vspace{-1.3cm}
\end{center}
\caption{\label{fig.2.} (Color online) Background substracted phonon spectra
divided by the Bose factor taken at $\xi=0.4$ (a),
$\xi=0.5$ (b), $\xi=0.6$ (c), $\xi=0.7$ (d), $\xi=0.8$ (e), and
$\xi=1$ (f) at $T=$ 30~K (black squares) and 300~K (red circles).}
\end{figure}
When phonons weakly couple to the conduction electrons their
energies and lifetimes slightly harden and narrow upon
cooling due to decreased anharmonicity. However, when phonons
strongly couple to conduction electrons in the nearly nested
regions of the Fermi surface, they soften and broaden upon cooling
as the Fermi surface sharpens. Inelastic neutron- or x-ray-scattering
measurements of phonons are the most
direct way to identify the phonon modes strongly coupled to
electrons. Due to the relatively large size of the available
single crystal, we carried out our investigation using inelastic
neutron scattering

Figure~1 illustrates how we determined the background that was
subtracted from the raw data to obtain the phonon  spectra. It shows
energy scans near the zone center at $T=$ 30~K. The strongest
peaks result from upward-dispersing acoustic phonons. The
inelastic peaks between 15 and 18~meV originate from
downward-dispersing optical phonons. In addition to phonons,
crystal-electric-field (CEF) transitions are present between 5
and 13~meV \cite{kreyssig04-2}. For the purpose of studying the
phonons, the CEF excitations contribute to the background. CEF
excitations are dispersionless, and their form factor is nearly
Q-independent in the narrow Q-range of interest. We determined
them from the spectra where they do not overlap with phonons.
The bold solid line in Fig. 1 represents the background, which includes
Gaussian peaks due to CEF excitations and a straight line due to
other energy-independent sources of background. It was subtracted
from the data to isolate the one-phonon scattering shown in
subsequent figures.

\begin{figure}\vspace{-5mm}
\begin{center}
\scalebox{0.45}{\includegraphics{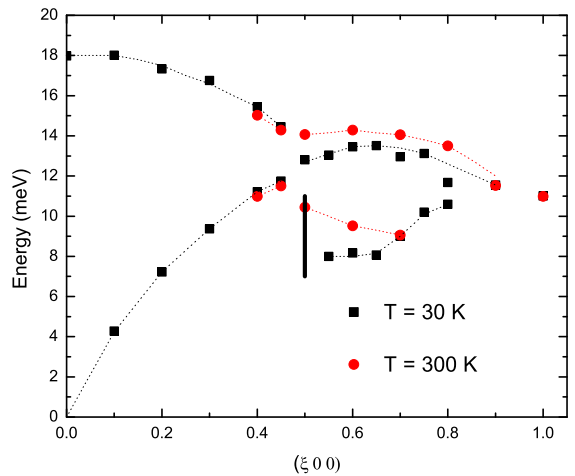}}\vspace{-1cm}
\end{center}
\caption{\label{fig.3.} (Color online) Low-energy phonons along
$[\xi ~0 ~0]$ at 30 and 300~K. The dashed lines are guides
for the eye. The vertical black bar at $\xi=$0.5 denotes a broad phonon
peak at 30~K containing also contributions from the next upper phonon
branch.}
\end{figure}

Figure ~2 shows the temperature dependencies of phonon spectra at
different wave-vector transfers ($Q$) to the neutron at 30 and 300~K.  
The CEF contribution was
subtracted from all low-temperature spectra as described above.
We did not attempt to identify the contribution of the CEF
excitations at 300~K, because they become weaker at high $T$,
whereas phonons become stronger due to the Bose factor, $n=1/(1-e^{-\hbar\omega/k_B T})$. 
Thus their contribution to the scattering intensity becomes
negligible. The data were divided by the Bose factor to correct
for the temperature dependence of the one-phonon scattering, which
also removed most temperature dependence from the background.
High-temperature curves were scaled with the low-temperature ones
by subtracting a constant, which is reasonable, because the
background increases with $T$. Phonon spectra of the optical
mode between 10 and 15~meV soften and broaden from 300 to 30~K at $Q=(0.5,0,8)$,
$Q=(0.6,0,8)$, $Q=(0.7,0,8)$, and $Q=(0.8,0,8)$, whereas the
spectra at $Q=(0.4,0,8)$, $Q=(0.9,0,8)$ (not shown), and
$Q=(1,0,8)$ are nearly temperature independent.

\begin{figure}\vspace{-5mm}
\begin{center}
\scalebox{0.4}{\includegraphics{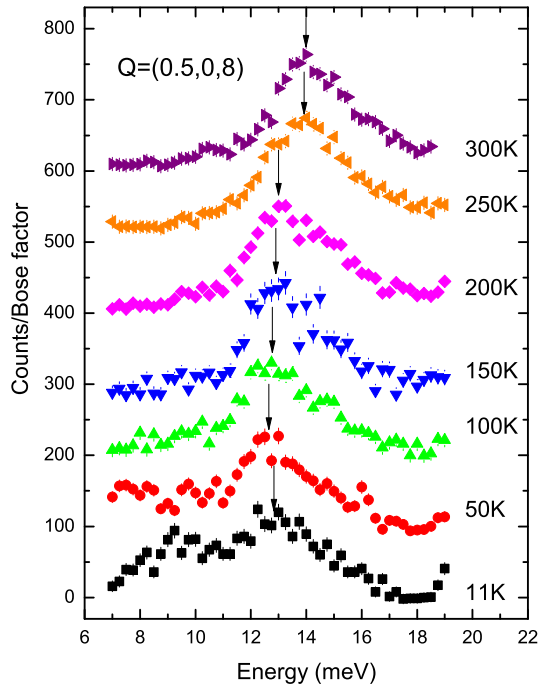}}\vspace{-1.3cm}
\end{center}
\caption{\label{fig.4.} (Color online) Phonon spectra divided by the Bose
occupation factor for a wave vector $\left(0.5,0,8\right)$ in the
temperature range from 11 to 300~K. The curves are vertically
shifted for clarity. The phonon-peak position is indicated by
arrows.}
\end{figure}

Figure~3 shows the dispersion of the two
low-energy acoustic and optical phonon branches propagating along
the $\left[\xi~0~0\right]$ direction at $T = $ 30 and 300~K. The
energy scans were fitted by Gaussian peaks. The dispersion of the
two interesting phonon branches at $T = $ 2~K is similar to the
dispersion measured in the temperature range $T = $ 11 --- 100~K
and there is no softening below 100~K \cite{kreyssig04}. However, strong
softening of both branches appears between 100 and 300~K between
$Q=$ (0.5,0,8) and (0.8,0,8). Our low-temperature data reveal a
discrepancy in the energy range $\hbar\omega=8-10.5$~meV as
compared to Ref. \cite{kreyssig04} which can be attributed to a
different interpretation of CEF contribution. Unlike Ref.
\cite{kreyssig04}, we observe a strong dip of the low-energy
phonon branch for wave vectors between $Q=$ (0.5,0,8) and
(0.75,0,8) at low temperatures, which is weaker but of the same
order of magnitude as the one found in superconducting
HoNi$_{2}$B$_{2}$C. This result indicates that
electron-phonon coupling in TbNi$_{2}$B$_{2}$C is strong enough
to mediate superconductivity (perhaps with a lower $T_c$) in the
absence of a pair-breaking mechanism.

Figure~4 illustrates the temperature dependence of the optic phonon
at Q=(0.5,0,8). Here we show the raw data divided by the Bose
factor without subtracting the background. The curves are
vertically shifted with the arrows indicating the phonon peak
position. The lowest temperature scan contains an additional
feature at 9~meV to which both the acoustic phonon and the
CEF excitation contribute. We have not measured the
temperature dependence of the CEF excitation, so it
is not possible to determine the temperature dependence of the
acoustic phonon based on the available data. However, the optic
phonon does not overlap with the CEF excitation and the data on
its \textit{T} dependence are unambiguous. We observe a softening from room
temperature to $T \cong $ 100~K of 1~meV. Below $\approx$ 100~K, it
shows no peak position shift.

\section{DISCUSSION AND CONCLUSIONS}
\label{sec:DISCUSSIONANDCONCLUSIONS}

Strong softening of two low-energy phonon branches was observed
by neutron scattering in superconducting
$\textit{R}$Ni$_{2}$B$_{2}$C single crystals with $R=$ Lu, Y, and
Er \cite{PhysRevB.52.R9839, PhysRevB.55.R8678,
PhysRevLett.77.4628, PhysRevB.66.212503}. Point-contact
spectroscopy revealed strong electron-phonon interaction in the
superconducting $\textit{R}$Ni$_{2}$B$_{2}$C compounds with $R=$
Y and Ho, in contrast to the nonsuperconducting
LaNi$_{2}$B$_{2}$C \cite{PhysRevLett.78.935}, which suggests that 
the presence of superconductivity is controlled by the strength of 
electron-phonon coupling as opposed to magnetic pair breaking.
However, our inelastic neutron scattering experiments on nonsuperconducting
TbNi$_{2}$B$_{2}$C clearly demonstrate strong electron-phonon
coupling in this compound. Thus we conclude that superconductivity in 
TbNi$_2$B$_2$C is absent due to magnetic pair breaking.

Previous studies of the competition between magnetism and
superconductivity in $\textit{R}$Ni$_{2}$B$_{2}$C superconductors with the
magnetic rare-earth elements $R=$ Lu, Tm, and Er revealed
a very weak coupling between the rare-earth magnetic moments and the
conduction electrons due to a small conduction-electron density at the
rare-earth site \cite{PhysRevB.50.647}. In contrast, a strong magnetic
pair-breaking effect has been observed in Dy and Tb samples \cite{PhysRevB.50.647}
which gives additional evidence that superconductivity in magnetic TbNi$_{2}$B$_{2}$C is indeed destroyed by the substantially strong interaction between the local magnetic moments and the conduction electrons. In summary, electron-phonon coupling in magnetic nonsuperconducting TbNi$_{2}$B$_{2}$C is strong enough to mediate superconductivity. This implies that its absence can only
result from magnetic pair breaking.

\section{ACKNOWLEDGMENTS}
\label{sec:ACKNOWLEDGMENTS}

The work by AK at the Ames Laboratory was supported by the U.S. Department of Energy,
Office of Basic Energy Science, Division of
Materials Sciences and Engineering, through Contract No. DE-AC02-07CH11358.

\bibliography{references}

\end{document}